
\input phyzzx
\input epsf

%
\ifx\epsfbox\UnDeFiNeD\message{(NO epsf.tex, FIGURES WILL BE
IGNORED)}
\def\figin#1{\vskip2in}
\else\message{(FIGURES WILL BE INCLUDED)}\def\figin#1{#1}\fi
\def\ifig#1#2#3{\xdef#1{fig.~\the\figno}
\goodbreak\midinsert\figin{\centerline{#3}}%
\smallskip\centerline{\vbox{\baselineskip12pt
\advance\hsize by -1truein\noindent\footnotefont{\bf
Fig.~\the\figno:} #2}}
\bigskip\endinsert\global\advance\figno by1}

 \hsize=15.8cm
\vsize=23cm
\voffset=0pt

\def\footnotefont{\tenpoint}

\newwrite\ffile\global\newcount\figno \global\figno=1
\def\fig{fig.~\the\figno\nfig}
\def\nfig#1{\xdef#1{fig.~\the\figno}%
\writedef{#1\leftbracket fig.\noexpand~\the\figno}%
\ifnum\figno=1\immediate\openout\ffile=figs.tmp\fi\chardef\wfile=
\ffile%
\immediate\write\ffile{\noexpand\medskip\noexpand\item{Fig.\
\the\figno. }
\reflabeL{#1\hskip.55in}\pctsign}\global\advance\figno by1\findarg}

\parindent 25pt
\overfullrule=0pt
\tolerance=10000

\nopagenumbers
\baselineskip=14pt

\line{\hfill CERN-TH/95-78}

\vskip 5cm
\centerline{A GAS OF D-INSTANTONS}
\vskip 1cm
 \centerline{ Michael B.  Green,\foot{ Permanent address: DAMTP,
Silver Street,
Cambridge CB3 9EW, UK\hfil\break\noindent email:
M.B.Green@amtp.cam.ac.uk}}
\centerline{Theory Division, CERN,}
\centerline{CH-1211, Geneva 23, Switzerland}

\vskip 2.0cm
\centerline{ABSTRACT}

A D-instanton is a space-time event associated with world-sheet
boundaries that contributes non-perturbative effects of order
$e^{-const/\kappa}$  to closed-string amplitudes.
Some properties of a gas of D-instantons are discussed in this paper.

\vskip 7.3cm
CERN-TH/95-78\hfil\break\indent
March 1995

\vfill\eject
\pagenumbers
\pageno=1
\sequentialequations

\REF\greena{M.B.  Green, {\it Space-Time Duality and Dirichlet String
Theory},
Phys.  Lett. {\bf 266B} (1991) 325.}Properties of string theory are
greatly
affected by world-sheet boundary conditions.  For example, a theory
may be
defined  by summing over world-sheets with boundaries on which the
string
space-time coordinates are required to satisfy constant Dirichlet
conditions --
the entire boundary is mapped to a point in the target space-time and
the
position of that point is then integrated, which restores
target-space
translation invariance ([\greena] and references therein).    The
result is a
theory that describes closed strings which possess dynamical
point-like
substructure as is indicated by the fact that fixed-angle scattering
is power
behaved as a function of energy.\REF\polchina{J.  Polchinski, {\it
Combinatorics
of World-Sheet Boundaries}, Phys.Rev. {\bf D50} (1994)
6041.}\REF\polchinb{ J.
Dai, R.G. Leigh and J. Polchinski {  \it New connections between
string
theories}, Mod. Phys. Lett.  {\bf A4} (1989) 2073.}\ Recently an
interesting
variation of this  scheme has been suggested  [\polchina]  (based on
[\polchinb]), involving the idea of \lq D-instantons'.  As the name
suggests,
these are world-sheet configurations which correspond to a target
space-time \lq
event', giving rise to exponentially suppressed contributions to
scattering
amplitudes behaving as  $e^{-C/\kappa}$ (where $\kappa$ is the
closed-string
coupling constant that is determined by the dilaton expectation value
and $C$ is
a constant).\REF\shenkera{S.H.  Shenker, {\it The strength of
nonperturbative
effects in string theory}, Proceedings of the Cargese Workshop on
Random
Surfaces, Quantum Gravity and Strings, Cargese, France (May 28 - Jun
1, 1990).}\
This is in accord with general observations in [\shenkera] that
suggest that
whereas instanton effects in  field theory typically behave as
$e^{-const./\kappa^2}$, analogous effects in closed-string theory
should behave
as
$e^{-C/\kappa}$.    In this paper the single D-instanton contribution
 will be
reexpressed as an exponential of an instanton \lq action'  and a gas
of  such
instantons will be defined.  The novel divergences associated with
Dirichlet
boundaries will be shown to cancel to all orders (as suggested in
[\polchina]).
The leading contribution to the free energy comes from free
D-instantons and is
of
order $\kappa^{-1}$  but corrections due to
long-distance interactions between D-instantons will be seen to be of
order
$\kappa^0$.

A general oriented string world-sheet has an arbitrary number of
boundaries and
handles.   In conventional open-string theories the embedding
coordinates
satisfy Neumann conditions on the boundaries while in the Dirichlet
case each
boundary is fixed at a space-time point, $y_B^\mu$ (where $B$ labels
the
boundary), which is then integrated.
The boundaries of moduli space are of various types.  There are the
usual
degenerations of cylindrical segments  that correspond to the
propagation of
physical closed-string states:
\item {(a)} Degeneration of handles.
\item{(b)} Degeneration of trivial homology cycles that divide a
world-sheet
into two disconnected pieces.

 \noindent In addition,  in the presence of boundaries there are
degenerations
of the following kinds:
\item{(c)}  A boundary may shrink to zero length giving the
singularities
associated with closed-string scalar states coupling to the vacuum
through the
boundary.
\item{(d)} Degeneration of strips forming open-string loops.  If both
string
endpoints are fixed at the same target-space point this gives an
infinite
contribution.
\item{(e)} Degeneration of trivial open-string channels, in which the
world-sheet divides into two disconnected pieces.  In the case of
Dirichlet
conditions the intermediate open string necessarily has both
end-points fixed at
the same space-time point leading to another infinite contribution.

The cohomology of the states of the open-string sector where the
string
end-points are fixed at $y_1^\mu$ and $y_2^\mu$ is isomorphic to that
of the
usual Neumann open string with momentum $p^\mu =  \Delta^\mu \equiv
y_2^\mu -
y_1^\mu$.  This may be viewed as a simple consequence of target-space
duality
and it means that the arbitrary diagram possesses a rich spectrum of
space-time
singularities, just as the usual loop diagrams possess a rich
momentum-space
singularity structure.  However, it is important to realize that the
wave
functions of these states depend on the mean position, $y^\mu =
(y_1^\mu +
y_2^\mu)/2$, in addition to $\Delta^\mu$ -- this extra variable has
no analogue
for
the usual open strings.  The intermediate open string in the trivial
degeneration (e) has $\Delta=0$ so its cohomology is isomorphic to
that of the
usual Neumann open-string theory when $p^\mu=0$. There is only one
physical
state in this case, which is the level-one vector.  This is the
isolated
zero-momentum physical state with a constant wave function in the
usual theory.
 However, in the Dirichlet theory its wave function $\zeta^\mu(y)$
is an
arbitrary function that is physical without the need to impose any
constraints
on it  -- it is a target-space Lagrange multiplier field.  The
presence of this
as an  intermediate state in a string diagram leads to a divergence
(the
propagator for the level-one state is singular since a Lagrange
multiplier field
has no kinetic terms).  The vertex operator that describes the
coupling of this
level-one vector state to a boundary is given by
$$g\oint d\sigma_B \zeta\cdot \partial_n X(\sigma_B,\tau_B) = i g
\zeta^\mu
{\partial\over \partial y_B^\mu},\eqn\veccoup$$
where  $\tau_B$ is the world-sheet position of the boundary and it is
fixed at
$y_B$ in the target space (and $g$ is the open-string coupling
constant that is
proportional to $\sqrt \kappa$).

At present there are two schemes for dealing with this level-one
divergence.  In
one of these the Lagrange multiplier field is eliminated by
integrating it,
thereby imposing a constraint  before the perturbation expansion of
the theory
is considered.  Some consequences of the presence of this constraint
were
discussed   in\REF\greenb{M.B. Green, {\it The Influence of
World-Sheet
Boundaries on Critical Closed String Theory}, Phys. Lett. {\bf B302}
(1993)
29.}\  [\greenb].  The other scheme [\polchina] uses combinatorics
for the sum
over boundaries that  is different from that of [\greena].  In this
case  the
divergences due to the level-one open-string field should cancel
between an
infinite number of diagrams as indicated in [\polchina].  The
divergences will
be shown to cancel in general in the formulation of the multi
D-instanton gas to
be be presented below.

As a preliminary, we shall obtain an equation for the contribution of
the
level-one divergences to a particular sum over connected (orientable)
world-sheets with Dirichlet boundaries. Consider a connected
orientable world-sheet  with $p_i$ boundaries fixed at any one of a
finite number of points $y_i$ (where
$i=1, \dots, n$ and $p_i =0, \dots, \infty$).  The string free
energy,
$$f_{p_1,p_2, \dots,p_n}(y_1,y_2,\dots,y_n),\eqn\freeen$$
is given by the usual multi-dimensional integral over the moduli
space of the
surface which has a total of $\sum_i p_i$ boundaries.  We shall be
interested in the sum over surfaces with all possible numbers of
boundaries for a given value of $n$,
$$S^{(n)} = \sum_{p_1,p_2,\dots,p_n=0}^\infty {1\over p_1! p_2! \dots
p_n!}
f_{p_1,p_2, \dots,p_n}(y_1,y_2,\dots,y_n), \eqn\facts$$
where the explicit combinatorial factor accounts for the symmetry
under the
interchange of identical boundaries  -- boundaries that are fixed at
the same
space-time point.  The definition of $f_{p_1,\dots, p_n}$ implicitly
contains a
sum over handles and the term with all $p_i=0$ in \facts\ is just the
usual
closed-string free energy, $S^{(0)}$.   Both types of open-string
degenerations,
(d) and (e),  described earlier lead to divergences in \facts\ due to
the
intermediate level-one open-string states and in each case the
coefficient of
the divergent term is proportional to the product of level-one vertex
operators
attached to the boundary at either end of the degenerating strip.  It
is
sufficient to consider single degenerations since the multiple
degenerations are
a subspace of these. The divergences of interest have the form
$\int_\epsilon^1
dq/q =\ln \epsilon$ where  $\epsilon$ is a world-sheet regulator.
Degenerations of type (e) divide the world-sheet into two factors, so
that if
the degenerating boundary is fixed at $y_1^\mu$ the singular term has
the form,
$$\eqalign{&\sum_{p_1,p_2,\dots, p_n;q_1,q_2,\dots,
q_n}{f_{p_1+q_1+1,p_2+q_2,
\dots,p_n+q_n} \over (p_1+q_1+1)! (p_2+q_2)! \dots (p_n+q_n)!}\sim
\cr
&  \ln \epsilon \left( {\partial \over \partial y_1^\mu}
\sum_{p_1,p_2,\dots
,p_n} {f_{p_1,p_2,\dots,p_n} \over p_1! p_2!\dots p_n!} \right)
\left(
{\partial \over \partial y_{1\mu} } \sum_{q_1,q_2,\dots, q_n}
{f_{q_1,q_2,\dots,q_n}\over q_1! q_2! \dots q_n!}
\right),\cr}\eqn\edegen$$
where the derivatives arise from two level-one  open-string vertex
operators
attached to two different boundaries.
A divergence also arises when an internal open-string with both ends
fixed at
the same point degenerates (this is a degeneration of type (d)).  In
this case
the coefficient of the divergence is proportional to two vertex
operators of the
level-one state attached to the same boundary giving,
$$\eqalign{& \sum_{p_1,p_2,\dots, p_n;q_1,q_2,\dots,
q_n}{f_{p_1+q_1+1,p_2+q_2,
\dots,p_n+q_n} \over (p_1+q_1+1)! (p_2+q_2)! \dots (p_n+q_n)!}\sim
\cr &
\ln \epsilon {\partial^2\over \partial  y_1^2} \sum_{p_1+q_1,
p_2+q_2,\dots,
p_n+q_n} {f_{p_1+q_1, p_2+q_2,\dots, p_n+q_n} \over (p_1+q_1)!
(p_2+q_2)! \dots
(p_n+q_n)!} .
\cr}\eqn\ddegen$$

\ifig\finsttwo{  The shaded blobs indicate a sum over world-sheets
with
boundaries inserted at any of the $n$ positions, $y_i$
($i=1,\dots,n$).  The
degeneration of a strip is indicated by a dashed line which either
divides the
surface into two (case (e)) or represents the degeneration of an
internal
open-string propagator (case (d)).  This gives a divergence that is
proportional
to the product of the total momenta entering the boundaries at either
end of the
line. }
{\epsfbox{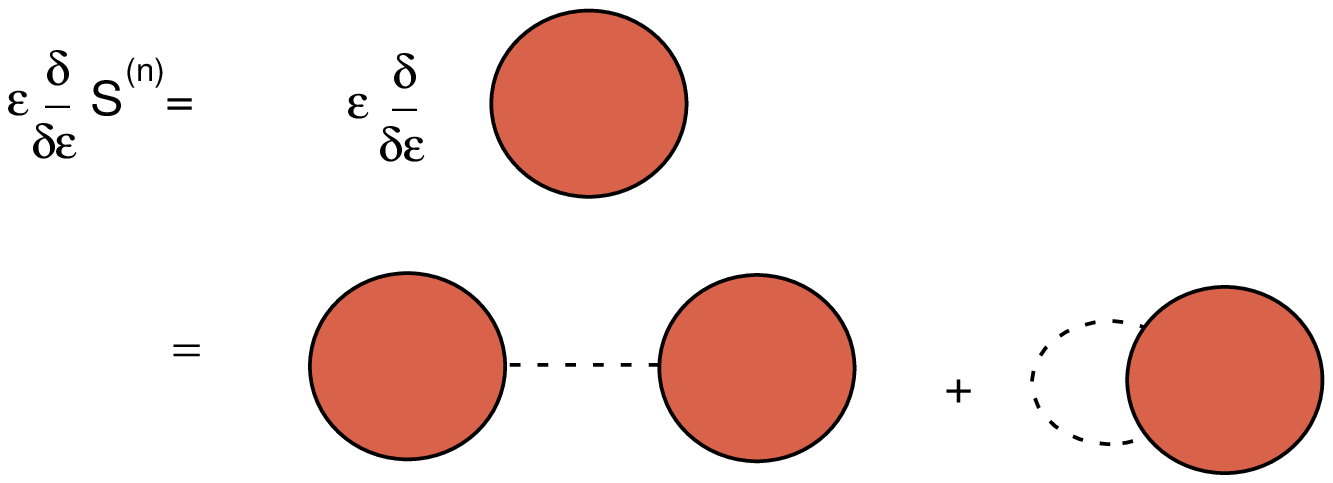}}
Combining \edegen\ and \ddegen\ and taking a derivative of the free
energy with
respect to $\epsilon$  extracts the dependence on the divergent
degenerations as
illustrated in   diagramatic form in \finsttwo,
$$ \epsilon {\partial \over \partial \epsilon}  S^{(n)} = \left(
{\partial \over
\partial y_1^\mu} S^{(n)} \right) \left( {\partial \over \partial
y_{1\mu}
}S^{(n)} \right)   + {\partial^2\over \partial  y_1^2} S^{(n)}
.\eqn\degenone$$
This equation is reminiscent of the renormalization group equation
expressing
the effect of  closed-string divergences in \REF\alwisa{S. de Alwis
and R.
Brustein, {\it Renormalization group equation and non-perturbative
effects in
string-field theory}, Nucl.  Phys.  {\bf B352} (1991) 451.}[\alwisa].

The rules for constructing the string partition function in the
presence of a
single D-instanton may be abstracted from the rules given in
[\polchina]  as
follows.  Firstly,
sum over world-sheets with insertions of any number of handles and
Dirichlet
boundaries that are all at the {\it same point} in the target space,
$y_1^\mu$,
which is to be integrated over.   The sum is now taken to include
{\it
disconnected} world-sheets although these do not appear to be
disconnected from
the point of view of the target space since the boundaries all meet
at the same
point.  A suitable symmetry factor  is to be included to take account
of
symmetry under the interchange of identical disconnected
world-sheets.  The
resulting one D-instanton partition function be expressed in
exponential form
as
$$Z^{(1)} = \int d^Dy e^{S^{(1)}(y)},\eqn\partfuh$$
where $S^{(1)}(y) = S^{(0)} + \sum_{p=1}^\infty f_p(y)/p!$ is now
interpreted as the one
D-instanton \lq action'
that is given by the $n=1$ term in \facts.  Scattering amplitudes may
be
generated from this expression if   $S^{(1)}$ is taken to be a
functional of the
background fields.

The requirement of consistent clustering properties in the target
space (as well as on the world-sheet) motivates
the following generalization that includes the sum over an arbitrary
number of
D-instantons (and which should be equivalent to the rather schematic
generalization motivated by  duality in [\polchina]).  This
involves summing over insertions of boundaries at any number of
positions,
$y_i$, that are to be integrated. The partition function is given in
the
language of a conventional instanton gas by the expression
$$Z = \sum_n {1\over n!}\left( \prod_{i=1}^n d^D y_i^\mu \right) e^{
S^{(n)}(y_1, \dots, y_n)},\eqn\partsuma$$
where $S^{(n)}$ is given by \facts\ and is now interpreted as the
action for $n$
interacting D-instantons.  Recall that  $S^{(n)}$ is defined by a
functional
integral over connected world-sheets and includes a term with no
boundaries
which is equal to $S^{(0)}$, the usual closed-string free energy.
The
expression for $Z$ therefore has the form,
$$Z= e^{S^{(0)}}\sum_n {1\over n!}\left( \prod_{i=1}^n d^D y_i^\mu
\right) e^{
S^{(n)\prime}(y_1, \dots, y_n)},\eqn\partsumb $$
where $S^{(n)\prime}$ is defined to be $S^{(n)}$ with the
zero-boundary term
missing.  In the general term in the sum any  boundary may be located
at any one
of the $n$ target-space positions, $y_i^\mu$, which are analogous to
the
collective
coordinates describing the positions of instantons in quantum field
theory.   It
is convenient to decompose $S^{(n) \prime}$ into those terms in which
all
boundaries are fixed at the same point (the free D-instanton terms),
those at which the boundaries are fixed at two different points
(two-body D-instanton interaction terms), those involving
three points (three-body D-instanton interactions), and so on,
$$S^{(n)\prime}(y_1, \dots, y_n) = \sum_{i=1}^n R_1 (y_i) +
\sum_{i\ne j}^n R_2
(y_i,y_j)+ \sum_{i\ne j\ne k}^n R_3 (y_i,y_j,y_k) + \dots
.\eqn\instact$$
It is important in writing this series to recall that the definition
of
$S^{(n)\prime}$ includes terms in which any subset of the $p_i$ are
zero --
these are terms that also contribute to the definition of
$S^{(m)\prime}$ with
$m<n$.
\ifig\finstone{ Contributions to the $n$ D-instanton action from the
first few
powers of $\kappa$.  a)  Diagrams with boundaries fixed at a single
space-time
point contribute to the free action.  b)  Diagrams with boundaries
fixed at two
points (indicated by the full and dashed boundaries) contribute to
the
two-instanton interaction.  c)  Diagrams contributing to the
three-instanton
interaction (at points indicated by full, dashed and dotted
boundaries).  The
coefficients explicitly show the combinatorical factors that arise
from symmetry
under the interchange of identical boundaries.
}
{\epsfbox{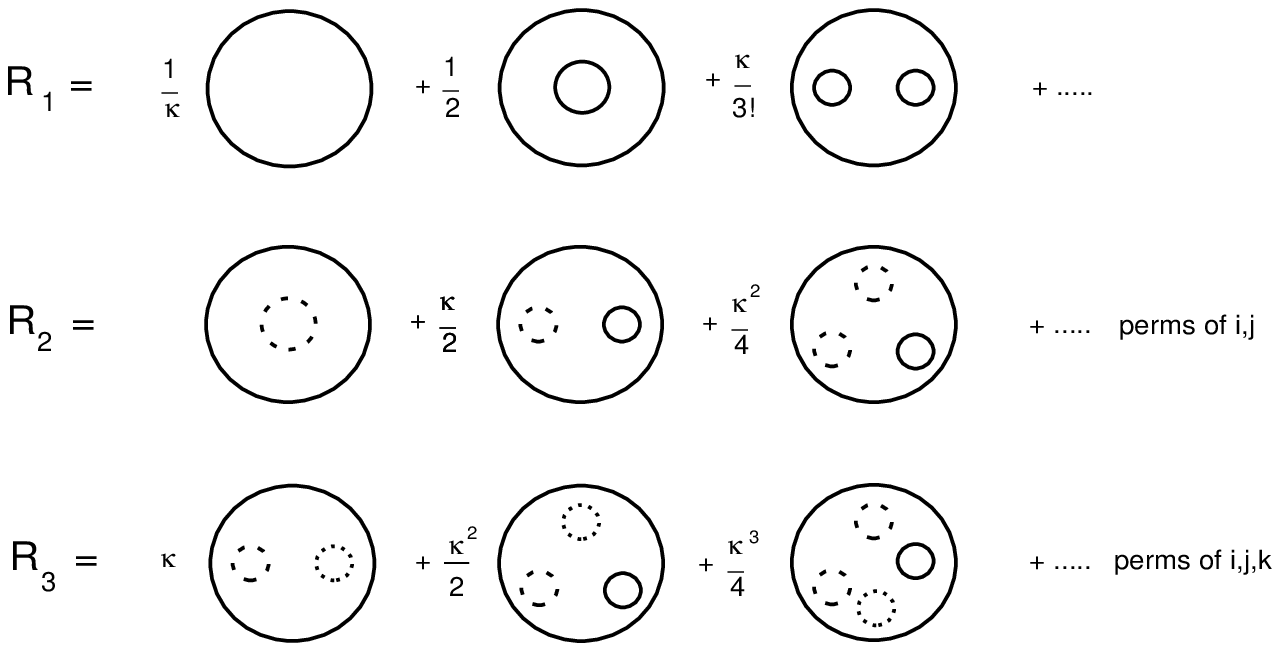}}

The free term in \instact, $R_1(y_i)\equiv S^{(1)\prime}(y_i)$, is
simply
given by the sum over connected orientable world-sheets of arbitrary
topology
with all boundaries fixed at a single point, $y_i$, illustrated in
\finstone
(a).  It is independent of $y_i$ (by translational invariance) and
has the form,
$$R_1=  -{C\over \kappa} + \ln D + O(\kappa),\eqn\leadterm$$
and thus $\sum_i R_1(y_i) = n R_1$.   The term in this series with
constant
coefficient, $C$, is determined by  functional integration over the
disk.
Explicit calculations determine $C=  2^8 \pi^{25/2} \alpha^{\prime
6}$ \REF\grinb{M. R. Douglas and B. Grinstein, {\it Dilaton tadpole
for the open bosonic string}, Phys. Lett. {\bf B183} (1987) 52; {\bf
B187} (1987) 442 (E).}
\REF\polchinc{J. Liu and J. Polchinski, {\it Renormalization of the
M\"obius
volume},  Phys. Lett. {\bf B 203}  (1988) 39.}  [\grinb,\polchinc].
It is somewhat remarkable that $C$ is a finite (positive) constant
with a value
that is consistent with the non-vanishing value of the disk with a
zero-momentum
dilaton attached.  Naively $C$ would be expected to vanish since it
should be
proportional to the inverse of the volume of the conformal Killing
group,
$SL(2,R)$ (which is infinite) but that would not be consistent with
the disk
with a zero-momentum dilaton insertion.   The $\zeta$ function
regularization in
[\grinb] and the proper-distance regularization in [\polchinc] lead
to the
subtraction of an infinite constant from the volume of the Killing
group, giving
the finite positive renormalized value of $C$.  The
$\kappa$-independent
constant $D$ in
\leadterm\ comes from the world-sheet annulus with both boundaries at
$y^\mu$.

Equation \leadterm\ leads to $e^{-C/\kappa}$ contributions to the
partition
function and to scattering amplitudes.  This has the qualitative form
expected
for non-perturbative effects in string theory on the basis of  matrix
models and
from the analysis of the rate of divergence of closed-string
perturbation theory
[\shenkera].  It is to be contrasted with a  characteristic feature
of
non-perturbative effects (such as instantons and solitons) in field
theory,
which behave as $e^{-const./\kappa^2}$.  This distinction between the
non-perturbative behaviour of quantum field theory and that expected
in
closed-string theory seems likely to be of great significance (some
possible
consequences are described in \REF\banksa{T.  Banks and M. Dine, {\it
Coping with
strongly coupled string theory},   RU-50-94, May 1994.}  [\banksa]).

The two-instanton interactions are given by the series of terms in
$R_2$  shown
in \finstone(b).  The diagrams contributing to $R_2$ are those in
which at least one boundary is fixed at either of the two space-time
points.  The leading terms in
this series have the form
$$R_2(y_i,y_j) =  f_{1,1}(y_i,y_j)   + {\kappa\over 2} \left(
f_{2,1}(y_i,y_j)
 +  f_{1,2}(y_i,y_j) \right)+ O(\kappa^2),\eqn\twobody$$
where $f_{p_i,p_j }(y_i,y_j )$ indicates a term with all $p_r=0$
apart from
$p_i$ and $p_j$.   The two-boundary term  is given by the expression
$$ f_{1,1}(y_i,y_j)  = c \int_0^\infty d\tau e^{-\Delta_{ij}^2 /\tau}
e^{2\tau}
\prod_{n=1}^\infty (1-e^{-2\tau})^{-24},\eqn\twodiff$$
where $c$ is a constant and $\Delta_{ij} = y_2-y_1$.
This diverges at the endpoint $\tau\to \infty$ due to the presence of
a
closed-string tachyon state.  This is a familiar problem of the
bosonic theory
which we shall bypass by tranforming to momentum space and declaring
that at low
momenta (or large distance) only the massless dilaton singularity
survives so
that in the  long-distance limit $\Delta^2 \to \infty$
$$f_{1,1}(y_i,y_j) \sim   |y_i - y_j|^{ 2-D} . \eqn\longdist$$
This coulomb-like behaviour due to dilaton exchange is analogous to
the
long-distance force between two classical instantons (magnetic
monopoles) in the
three-dimensional euclidean Georgi--Glashow model.  However, unlike
the case of
magnetic monopoles the interaction term, \longdist\  is not of the
same order in
$\kappa$ as the leading term,  $S^{(1)\prime}$.

It is straightforward to show that the partition function defined by
\partsumb\ does not have the divergences arising from the level-one
open-string
vector state.  The term in the partition function coming from $n$
D-instantons
has
a dependence on $\epsilon$ that can be written by expanding the
exponent to
first order in $\ln \epsilon$ using \degenone, giving,
$$\eqalign{&\ln \epsilon \int \prod_{i=1}^n d^Dy_i^\mu \left\{\left(
{\partial
\over \partial y_1^\mu} S^{(n)} \right) \left( {\partial \over
\partial y_{1\mu}
}S^{(n)} \right)   + {\partial^2\over \partial  y_1^2}
S^{(n)}\right\}  e^{
S^{(n)}}
 = 0.\cr}\eqn\finitr$$
The fact that the expression vanishes makes use of an integration by
parts of
the second term.

The cancellation is illustrated in an example in fig.~3.  This shows
contributions to a particular divergence coming from the sum of (a) a
planar
connected world-sheet, (b) a planar disconnected world-sheet and (c)
a
non-planar disconnected world-sheet.    The sum of these
contributions may be
written symbolically as
$$(a)+ (b) + (c) \ = \  \ln \epsilon \int d^Dy_1^\mu{\partial^2\over
\partial
y_1^2}\left( f_{1,1}(y_1,y_2)
f_{2,1}(y_1,y_2)\right),\eqn\abcdegen$$
which vanishes after integration over $y_1$ (assuming suitable
boundary
conditions).
The cancellation of divergences evidently involves a conspiracy
between terms
with different numbers of boundaries and handles.  Therefore it is
only possible
when the
boundary weight has a specific value -- it is not possible to add
Chan--Paton
factors to the boundaries as is usually the case in open string
theories.
It is disturbing that the cancellation of the divergences requires an
integration by parts which looks nonlocal since the D-instanton gas
is supposed
to
satisfy clustering properties that express the locality of the
theory.  However,
(at least in flat space) the potentially dangerous surface terms that
arise are
suppressed since they involve interactions between boundaries fixed
at widely
separated points.
\ifig\finsthre{ An example of the cancelation of a divergence that
requires
world-sheets with handles. a) One of the divergent degenerations of a
world-sheet with two boundaries at $y_1$ (full lines) and two at
$y_2$ (dashed
lines).  b)  A degeneration of a disconnected planar  world-sheet
that, after
integration over $y_1$,  contributes to the same divergence as in a).
 c)   A
degeneration on a disconnected world-sheet with a handle that gives a
divergence
that adds to  the divergences in b) and c) to give a total $y_1$
derivative.}
{\epsfbox{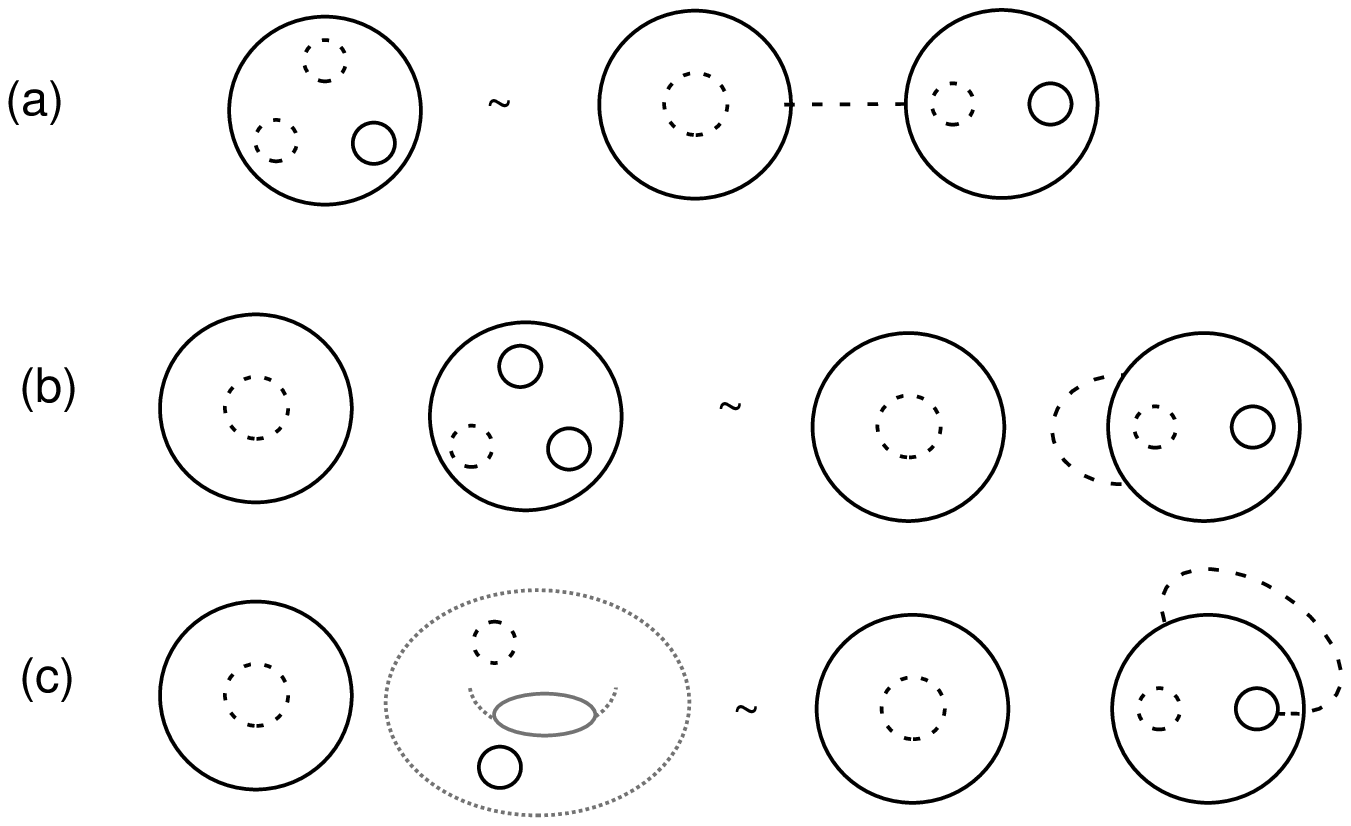}}

On-shell scattering amplitudes may be defined in the usual manner by
considering
fluctuations in the background fields, resulting in closed-string
vertex
operators coupled to the world-sheets.  The connected amplitudes are
obtained
from $\ln Z$, where the word \lq connected' refers to the target
space.  It is
straightforward to show,  closely following the usual field theory
combinatoric
arguments, that $\ln Z$ generates only amplitudes in which all vertex
operators
are attached to world-sheet components with at least one boundary
fixed at a
common target-space point.  These represent terms that are connected
in the
target space even though they involve disconnected world-sheets.  The
 terms
that are disconnected in the target space (terms in which vertex
operators are
attached to disconnected world-sheet components that do not have any
boundary
fixed at a common point) cancel out of the expansion of $\ln Z$.

 As an example, we may ask how light particle masses depend on the
presence of
the boundaries.  Thus, the mass of the dilaton is given by the
dilaton two-point
function evaluated at a minimum of the dilaton potential, assuming
there is one.
 The leading contribution from a single D-instanton comes from the
diagram in
which the two dilaton vertex
operators are inserted in disconnected disks with boundaries fixed at
a common
target-space point, $y^\mu$, which is integrated.  The result is of
the form  $
(\kappa)^0\  e^{-R_1/\kappa}$ where $R_1$ is given by \leadterm.
However, at this level of understanding the dilaton potential  does
not have a
minimum at a finite value of the dilaton field so that the value of
$\kappa$ is
incorrect and this term has no relation to the actual dilaton mass.

Now consider a process with two on-shell closed-string  second-rank
massless
tensor states with polarization tensors $\zeta^{(i)}_{\mu\nu}$ and
momenta
$p^{(i)\mu}$ ($i=1,2$)
satisfying the covariant conditions,
$$p^{(i)\mu} \zeta^{(i)}_{\mu\nu} = p^{(i)\mu} \zeta^{(i)}_{\nu\mu} =
0, \qquad
\zeta^{(i)\mu}_{\ \ \ \ \mu} =0, \qquad p^{(i)2} = 0. \eqn\polarcon$$
The lowest-order single D-instanton contribution with these external
states
comes from a diagram in
which both vertex operators are attached to the same disk with the
boundary
fixed at $y^\mu$.  Before integration over $y^\mu$ this defines a
form factor
that can be expressed as
$\kappa De^{-C/\kappa} e^{iq\cdot y}F(1,2;q)$ (where $q^\mu =
p^{(1)\mu} +
p^{(2)\mu}$) with $F(1,2;q)$ given by the remarkably simple gauge
invariant
expression \REF\greenwai{M.B.  Green  and P. Wai, {\it Inserting
Boundaries in
World-Sheets} with P.  Wai, Nucl.  Phys.  {\bf B431} (1994)
131.
}[\greenwai],
$$\eqalign{&   F(1,2;q) = \zeta^{(1)\mu\nu}
\zeta^{(2)}_{\mu\nu} {1\over q^2 - 8} + \zeta^{(1)\mu\nu}
\zeta^{(2)}_{\nu\mu}
{1\over q^2 + 8} - {1\over 2} \zeta^{(1)\nu\mu} \zeta^{(2)}_{\rho\nu}
q^\rho
q_\mu  \left[{1\over q^2} - {1\over q^2 + 8}\right]  \cr
&   +  {1\over 2}
\zeta^{(1)\mu\nu}\zeta^{(2)}_{\rho\nu} q^\rho q_\mu  \left[{1\over
q^2} -
{1\over q^2 - 8}\right] +  {1\over 16}
\zeta^{(1)}_{\mu\nu}\zeta^{(2)}_{\rho\sigma} q^\rho q^\sigma q^\mu
q^\nu
\left[{1\over q^2-8} - {2\over q^2} + {1 \over q^2+8} \right].
\cr}\eqn\gravivert$$
With external gravitons $\zeta^{(i) \mu\nu}$ is symmetric and the
result
vanishes  after the $y^\mu$ integration (which sets $q^\mu = 0$), so
the
graviton does not develop a mass.  The
terms with poles at $q^2=0$ arise from the coupling of the dilaton to
the vacuum
and would be absent if the vacuum state were defined properly as a
minimum of
the dilaton potential.

 The lowest-order contribution to the scattering amplitude with four
gravitons
comes from the diagram in which a pair of vertex operators are
coupled to one
world-sheet disk and another pair to a different disk where the
boundaries of
the two disks are fixed at the same point, $y^\mu$, that is
integrated.  The
amplitude is therefore given by
$$A = \kappa^2 D e^{-C/\kappa} \int d^Dy F(1,2;y) F(3,4;y)
,\eqn\fourgrav$$
which is proportional to the convolution of two form factors.

The inconsistencies of the critical bosonic string  make it difficult
to
interpret this theory in more detail. In particular, it is not at all
clear in
what sense these ideas  make contact with more conventional instanton
ideas,
such as those that arise in matrix models.   It would be of interest
to
study similar boundary effects in two-dimensional bosonic string
theories  and
compare them with other descriptions of instantons in such theories
(such
as\REF\ovruta{ R. Brustein, M. Faux, and B. A. Ovrut, {\it  Instanton
effects in
matrix models and string effective lagrangians},  CERN-TH-7301-94;
hep-th/9406179 }\ [\ovruta]). One peculiarity, at least in the
bosonic theory,
is that there are no anti D-instantons. \REF\greenm{M.B. Green, {\it
Point-like
states for type
2b superstrings}, Phys.  Lett. {\bf B329}  (1994) 435.}\   The
absence of the
tachyonic closed-string singularity in the
superstring also suggests that this may be an arena where a more
consistent
discussion could be given [\greenm].   The disk diagram within
superstring
theory is again a finite constant, given by the non-zero coupling of
the dilaton
to the disk with one particular spin structure of the world-sheet
fermions.

\vskip 0.5cm
\noindent{\it Acknowledgments}  \hfill\break
I am grateful to Ramy Brustein, Burt Ovrut, Michael Gutperle and Eric
Verlinde for useful
conversations.

\refout
\bye